\documentclass[notitlepage]{article}
\usepackage[margin = 1in]{geometry}
\usepackage[numbers, sort&compress]{natbib}
\usepackage{authblk}

\usepackage{amsmath, amsfonts, amssymb, amsthm, dsfont}
\usepackage{mathtools}
\usepackage{hyperref}
\usepackage{physics}
\usepackage{tabularx}
\usepackage{tikz}
\usepackage{usefulnotations}
\usepackage{pifont}

\begin{document}
	\title{\bf Evolution Generators for Complex Parameters}

	\author[1]{Jen-Yin Yeh}
	\author[1, 2, 3, 4, *]{Chia-Yi Ju}

	\affil[1]{Department of Physics, National Sun Yat-sen University, Kaohsiung 804201, Taiwan}
	\affil[2]{Center for Theoretical and Computational Physics, National Sun Yat-sen University, Kaohsiung 804201, Taiwan}
	\affil[3]{Research Center for Quantum Computing and Quantum Materials, National Sun Yat-sen University, Kaohsiung 804201, Taiwan}
	\affil[4]{Physics Division, National Center for Theoretical Sciences, Taipei 106319, Taiwan}

	\maketitle

	\let\thefootnote\relax
	\footnotetext{* chiayiju@mail.nsysu.edu.tw}

	\begin{abstract}
		Theoretical studies on how quantum systems are affected by external factors are often formulated through parameter changes in the system's Hamiltonian. Beyond Berry connections and phases, which focus on specific Hamiltonian eigenstates, recent studies inspired by non-Hermitian quantum formalisms provide a framework for obtaining general state evolution generators for real-valued parameters. This study extends the applicability of the evolution generator formalism to complex-valued parameters via Wirtinger derivatives. By treating a complex parameter and its conjugate as independent variables, the evolution equations are derived for both quantum states and the metric of the Hilbert space bundle. The analysis demonstrates that while state evolution with respect to a complex parameter is naturally governed by its corresponding evolution generator, metric evolution requires a coupled contribution from both the generator and its conjugate counterpart to preserve state normalization. Explicit examples are worked out to illustrate the implementation and physical consistency of the formalism.
	\end{abstract}

	\section{Introduction}

		In most physical systems, the evolution of a quantum system is governed by a Hamiltonian that depends explicitly on external control parameters, such as coupling constants, tunneling amplitudes, or driving field amplitudes~\cite{Ising1925, Heisenberg1928, Potts1952, Jaynes1963, Hubbard1963, Kondo1964, Su1979}. Tracking how a quantum state and its underlying properties respond to continuous variations of these parameters is essential, especially for understanding quantum phase transitions~\cite{Tu2022, Tzeng2023, Tu2023, Wu2023, Li2025, Tzeng2025, Znojil2026, Ju2026}, perturbation theory~\cite{Schroedinger1926, Kato1976, Caliceti1980, Znojil2020b, Znojil2024a, Chen2025}, or even phenomena around excpetional points in non-Hermitian systems~\cite{Heiss2004, Dembowski2004, Zhong2018, Kawabata2019, Oezdemir2019, Znojil2020a, Tzeng2021, Znojil2021, Wiersig2022, Hu2022, Loran2022, Wiersig2023, Arkhipov2023, Jia2023, Naikoo2023, Guria2024, Lai2024, Beniwal2024, Roy2025, Ince2025, Ju2025b}. Because these physical parameters are frequently complex-valued, developing a theoretical tool for capturing the response of quantum states to these parameters is essential.

		Many theoretical studies on the parameter-responses are through Berry connections and Berry phases~\cite{Born1928, Kato1950, Berry1984, Xiao2010}. These connections and phases are the response from a specific eigenstate or multiple eigenstates~\cite{Simon1983, Zak1989, Pachos1999, Chang2008, MehriDehnavi2008, Rhim2017, Palumbo2021, Tang2022, Guo2023, Ju2025a, Arkhipov2026}. Nevertheless, the eigenstates are just special cases of a general quantum state, which does not necessarily reflect the most general features of a quantum system.

		Recent studies~\cite{Ju2024, Ju2024a} show that, inspired by non-Hermitian quantum mechanics formalisms~\cite{Bender1998, Bender2002, Mostafazadeh2003, Mostafazadeh2004, Bender2004, Bender2007, Mostafazadeh2010, Brody2013, Ju2019, Ju2025}, by assuming that physical quantum states remain physical when varying the parameters, the connections, or the parameter evolution operators, of the full Hilbert space bundle of the corresponding quantum system can be obtained via a systematic method. Therefore, it is very useful for understanding how the full quantum system changes with the external parameters.

		In the framework, the parameters are restricted to be real numbers for mathematical consistency. However, the corresponding parameters of external changes to the system are not necessarily real. Of course, like many studies, the complex parameters can be decomposed into real ones. Nevertheless, this is a clear restriction to the framework.

		Consequently, this study extends the discussions of the real parameter evolutions for the full Hilbert space bundle to complex ones. To be more specific, this work starts by decomposing complex parameters and their conjugate parameters into real ones, and then recombining them to form a formalism for complex parameters.

	\section{Brief Review of the Emergent Parallel Transports \label{Sec:Review}}

		The following section summarizes the core framework from Ref.~\cite{Ju2024}, incorporating key refinements from Ref.~\cite{Ju2024a}. Comprehensive derivations and in-depth discussions can be found in the original studies.

		It is well-known that the time evolution operator is not unitary for a non-Hermitian Hamiltonian. Therefore, even for a state normalized at a given time-slice, time evolution does not maintain its normalization. To be more specific, if a quantum state $\ket{\Psi(t)}$ evolves under a non-Hermitian Hamiltonian,
		\begin{align}
			\braket{\Psi (t_0)}{\Psi (t_0)} = 1 \quad \not\Rightarrow \quad \braket{\Psi (t)}{\Psi (t)} = 1.
		\end{align}
		Consequently, the standard inner product in quantum mechanics does not preserve the probability for non-Hermitian systems.

		To resolve the issue, several studies~\cite{Bender1998, Bender2002, Mostafazadeh2003, Bender2004, Mostafazadeh2004, Bender2007, Mostafazadeh2010, Brody2013} have proposed modifying the inner product from the conventional $\braket{\Phi}{\Psi}$ to
		\begin{align}
			\Braket{\Phi}{\Psi} = \bra{\Phi} G \ket{\Psi},
		\end{align}
		where $\Bra{\Phi} = \bra{\Phi} G$, $\bra{\Phi} = \ket{\Phi}^\dagger$, and $\Ket{\Psi} = \ket{\Psi}$, such that the operator $G$ acts as a metric for the corresponding Hilbert space bundle. To ensure that this qualifies as a valid inner product, $G$ must satisfy specific properties: it must be positive-definite ($G > 0$) and Hermitian ($G = G^\dagger$).

		To systematically find the metric $G$, the Schr\"{o}dinger equation can be treated as a parallel transport equation~\cite{Mostafazadeh2018, Ju2019}, namely,
		\begin{align}
			0 = \nabla_t \ket{\Psi} \equiv \left(\partial_t + i H\right) \ket{\Psi}, \label{SchroedingerEq}
		\end{align}
		where $\nabla_t$ is the connection in the $t$-direction and $H$ is the Hamiltonian of the quantum system. For the metric $G$ to be compatible with this connection (i.e., with the Schro\"{o}dinger equation), it must satisfy
		\begin{align}
			0 = \nabla_t G = \partial_t G - \i G H + \i H^\dagger G. \label{MetricTEq}
		\end{align}

		Notably, this inner product reduces to the conventional one, $\Braket{\Phi}{\Psi} = \braket{\Phi}{\Psi}$, for systems with $G = \mathds{1}$. A closer look at Eq.~\eqref{MetricTEq} reveals that this condition holds exclusively for Hermitian systems, where $H = H^\dagger$.

		Furthermore, the metric $G$ preserves state normalization and consequently total probability, while also inherently depending on the Hamiltonian.

		Because $G$ depends on the Hamiltonian, any additional parameter $q$ in the Hamiltonian induces a corresponding $q$-dependence in $G$. A previous study~\cite{Ju2024} demonstrated that the $q$-derivative of the metric $G$ must formally satisfy
		\begin{align}
			\partial_q G = \i G K - \i K^\dagger G, \label{MetricQEq}
		\end{align}		
		owing to the positive-definiteness and Hermiticity of $G$ for $q \in \mathbb{R}$. Specifically, this form guarantees that
		\begin{align}
			\left(\partial_q G\right)^\dagger = \partial_{\bar{q}} G^\dagger = \partial_q G, \label{HermiticityPartialQG}
		\end{align}
		for $q = \bar{q}$. Furthermore, enforcing the preservation of normalization determines the $q$-dependence of the quantum states, which results in
		\begin{align}
			\partial_q \ket{\Psi} = - \i K \ket{\Psi}. \label{StateQEq}
		\end{align}

		In fact, due to the similarity between Eq.~\eqref{StateQEq} and the Schr\"{o}dinger equation, it is natural to rewrite Eqs.~\eqref{MetricQEq} and \eqref{StateQEq} in terms of covariant derivatives (or connections). Together with Eqs.~\eqref{MetricTEq} and \eqref{MetricQEq}, these relations form the set
		\begin{align}
			& \nabla_t \ket{\Psi} = \left(\partial_t + \i H\right) \ket{\Psi} = 0,\\
			& \nabla_q \ket{\Psi} = \left(\partial_q + \i K\right) \ket{\Psi} = 0,\\
			& \nabla_t G = \partial_t G - \i G H + \i H^\dagger G = 0,\\
			& \nabla_q G = \partial_q G - \i G K + \i K^\dagger G = 0.
		\end{align}
		Moreover, the compatibility condition among these four equations results in
		\begin{align}
			\partial_t K - \partial_q H + \i \left[H, K\right] = 0. \label{NonuniqueSingle}
		\end{align}

		Although Eq.~\eqref{NonuniqueSingle} is a differential equation and therefore does not uniquely determine $K$, alternative solutions for $K$ are related by a gauge transformation. Consequently, distinct choices of $K$ correspond to different gauge fixings, allowing $K$ to be freely chosen for convenience.

		This discussion naturally extends to the multi-parameter case. When the quantum system depends on a set of parameters $\{q^1, q^2, \dots, q^n\}$, the formulation expands accordingly, with superscripts denoting parameter indices rather than exponents.

		Repeating the preceding steps and introducing the shorthand notations $\nabla_i \equiv \nabla_{q^i}$ and $\partial_i \equiv \partial_{q^i}$ to avoid clutter (where no confusion arises), the evolution equations take the form
		\begin{align}
			& \nabla_i \ket{\Psi} = \left(\partial_i + \i K_i\right) \ket{\Psi} = 0,\\
			& \nabla_i G = \partial_i G - \i G K_i + \i K_i^\dagger G = 0,
		\end{align}
		along with the compatibility condition
		\begin{align}
			\partial_t K_i - \partial_i H + \i \left[H, K_i\right] = 0. \label{NonuniqueMultiple}
		\end{align}

		For the same reason mentioned in the single-parameter case, the operators $K_i$ are not uniquely determined by Eq.~\eqref{NonuniqueMultiple}. However, in the multi-parameter case, the gauges cannot be chosen independently; instead, they must satisfy the additional compatibility condition
		\begin{align}
			\partial_i K_j - \partial_j K_i + \i \left[K_i, K_j\right] = 0, \label{MultiparameterConstraint}
		\end{align}
		for all $i$ and $j$.

		For a more comprehensive review of this framework, interested readers are encouraged to consult Refs.~\cite{Ju2024, Ju2024a}.

		It is worth noting, however, that the preceding discussion relies on an assumption not explicitly mentioned in Refs.~\cite{Ju2024, Ju2024a}, namely that $q^i \in \mathbb{R}$. If any of the parameters $q^i$ are complex, Eq.~\eqref{MetricQEq} breaks down because the relation $\partial_{\bar{q}^i} G^\dagger = \partial_{q^i} G$ [the final equality in Eq.~\eqref{HermiticityPartialQG}] no longer holds.

		This work, on the other hand, aims to resolve this limitation. Because Hamiltonian components or coefficients are frequently complex, even in Hermitian systems, addressing this issue is essential to ensure the broad applicability of the framework.

	\section{Complex-Parameter Evolution Generator}

		In quantum mechanics, a system's state and evolution are fundamentally governed by its Hamiltonian. Consequently, physical changes or environmental variations manifest as modifications to the parameters within the Hamiltonian, such as coupling constants, energy splittings, or external field amplitudes. In many realistic scenarios, these parameters are naturally represented as complex numbers. Therefore, extending the parallel transport framework from real parameters to complex parameters is essential for broader physical applicability.

		To address this, rather than modifying the expression in Eq.~\eqref{MetricQEq}, any complex number $z$ and its conjugate $\bar{z}$ can be decomposed into two real numbers $x$ and $y$ such that
		\begin{align}
		 & z = x + \i y, \\
			& \bar{z} = x - \i y.
		\end{align}
		The Wirtinger derivatives (specifically the $z$- and $\bar{z}$-derivatives) can then be expressed in terms of the $x$- and $y$-derivatives as
		\begin{align}
			& \partial_z = \dfrac{1}{2} \left(\partial_x - \i \partial_y\right),\\
			& \partial_{\bar{z}} = \dfrac{1}{2} \left(\partial_x + \i \partial_y\right). \label{WirtingerDerivative}
		\end{align}
		In other words, complex derivatives decompose naturally into real derivatives.

		Thus, if the Hamiltonian depends on a complex variable $z$ (and optionally its conjugate $\bar{z}$), denoted as $H = H(z)$ [or more generally, $H = H(z, \bar{z})$ or $H = H(t, z, \bar{z})$], it can be equivalently reformulated as a function of two real variables, e.g.,
		\begin{align}
			H (z, \bar{z}) = H (x + \i y, x - \i y).
		\end{align}
		Through this approach, the evolution generators in the $x$- and $y$-directions ($K_x$ and $K_y$) are obtained via
		\begin{align}
			& \partial_t K_x - \partial_x H + \i \left[H, K_x\right] = 0, \label{KXYEq1}\\
			& \partial_t K_y - \partial_y H + \i \left[H, K_y\right] = 0, \label{KXYEq2}\\
			& \partial_x K_y - \partial_y K_x + \i \left[K_x, K_y\right] = 0. \label{KXYEq3}
		\end{align}

		Employing the Wirtinger derivatives given in Eq.~\eqref{WirtingerDerivative}, the $z$-evolution equation for the quantum state can be decomposed into its $x$- and $y$-components, namely,
		\begin{align}
			\partial_z \ket{\Psi} & = \frac{1}{2} \left(\partial_x - \i \partial_y\right) \ket{\Psi} = - \frac{\i}{2} \left(K_x - \i K_y\right) \ket{\Psi} = - \i K_z \ket{\Psi},
		\end{align}
		where the effective generator $K_z$ is defined as
		\begin{align}
			K_z \equiv \frac{K_x - \i K_y}{2}.
		\end{align}
		Proceeding analogously for the conjugate coordinate $\bar{z}$, the corresponding evolution equation reads
		\begin{align}
			\partial_{\bar{z}} \ket{\Psi} = - \i K_{\bar{z}} \ket{\Psi},
		\end{align}
		with the generator $K_{\bar{z}}$ defined by
		\begin{align}
			K_{\bar{z}} \equiv \frac{K_x + \i K_y}{2}.
		\end{align}

		Attention now turns to the parameter evolution of the metric $G$. It is worth noting that even if the Hamiltonian depends only on $z$ and not on $\bar{z}$, the metric $G$ still depends on both parameters. One way to see this is via Eq.~\eqref{MetricTEq}, where the time evolution of $G$ depends on both $H$ and $H^\dagger$. Therefore, even if $H = H(z)$, the contribution from $H^\dagger$ introduces a dependence on $\bar{z}$.

		Consequently, the $z$- and $\bar{z}$-evolution equations for the metric $G$ are not generated by a single $K_z$ or $K_{\bar{z}}$, but rather by a mixture of the two, namely,
		\begin{align}
			& \partial_z G = \i G K_z - \i K_{\bar{z}}^\dagger G, \label{GZEq}\\
			& \partial_{\bar{z}} G = \i G K_{\bar{z}} - \i K_z^\dagger G. \label{GZBarEq}
		\end{align}
		Detailed derivations of these and subsequent relations can be found in Appendix~\ref{Appendix:DetailedDerivations}.

		Although it might appear somewhat peculiar that both $K_z$ and $K_{\bar{z}}$ appear in the $z$-evolution of the metric, this structure is strictly required to preserve state normalization. Specifically, the $z$-derivative of the dual state $\bra{\Psi}$ is governed by $K_{\bar{z}}$, rather than $K_z$, through Hermitian conjugation, namely,
		\begin{align}
			\left(\partial_{\bar{z}} \ket{\Psi}\right)^\dagger = \left(- \i K_{\bar{z}} \ket{\Psi}\right)^\dagger \quad \Rightarrow \quad \partial_z \bra{\Psi} = \i \bra{\Psi} K_{\bar{z}}^\dagger.
		\end{align}
		Consequently, the presence of $K_{\bar{z}}^\dagger$ in the $z$-derivative of the metric $G$ ensures that
		\begin{align}
			\partial_z \Bra{\Psi} = \partial_z \left(\bra{\Psi} G\right) = \i \Bra{\Psi} K_z,
		\end{align}
		which directly preserves the state normalization due to $\partial_z \Braket{\Psi}{\Psi} = 0$.

		An entirely analogous argument applies to the $\bar{z}$-evolution of the metric in Eq.~\eqref{GZBarEq}, where the appearance of $K_z^\dagger$ ensures the consistent preservation of the dual inner product under $\bar{z}$-variations.

		By repeatedly applying the Wirtinger derivatives from Eq.~\eqref{WirtingerDerivative} along with the relations in Eqs.~\eqref{KXYEq1}, \eqref{KXYEq2}, and \eqref{KXYEq3}, the governing equation for $K_z$ is found to be
		\begin{align}
			& \partial_t K_z - \partial_z H + \i \left[H, K_z\right] = 0,\label{KZEq}\\
			& \partial_t K_{\bar{z}} - \partial_{\bar{z}} H + \i \left[H, K_{\bar{z}}\right] = 0. \label{KZBarEq}
		\end{align}

		Therefore, the covaraint derivative version of the above equations can also be expressed as
		\begin{align}
			& \nabla_z \ket{\Psi} = \left(\partial_z + \i K_z\right) \ket{\Psi} = 0, \label{StateZEvolution}\\
			& \nabla_{\bar{z}} \ket{\Psi} = \left(\partial_{\bar{z}} + \i K_{\bar{z}}\right) \ket{\Psi} = 0, \label{StateZBarEvolution}\\
			& \nabla_z G = \partial_z G - \i G K_z + \i K_{\bar{z}}^\dagger G = 0,\\
			& \nabla_{\bar{z}} G = \partial_{\bar{z}} G - \i G K_{\bar{z}} + \i K_z^\dagger G = 0.
		\end{align}

		In addition to Eqs.~\eqref{KZEq} and \eqref{KZBarEq}, $K_z$ and $K_{\bar{z}}$ are not completely independent, but are instead related by
		\begin{align}
			\partial_z K_{\bar{z}} - \partial_{\bar{z}} K_z + \i \left[K_z, K_{\bar{z}}\right] = 0, \label{KZandKZBarEq}
		\end{align}
		which also ensures the consistency between Eqs.~\eqref{StateZEvolution} and \eqref{StateZBarEvolution}.

		Table~\ref{Table:Comparison} summarizes the evolution generators, state evolutions, and metric evolutions for both real parameters $\{q, q'\}$ and the complex conjugate pair $\{z, \bar{z}\}$.

		\begin{table}[t]
			\centering
			\begin{minipage}{0.9\textwidth}
				\renewcommand{\arraystretch}{1.3}
				\setlength{\tabcolsep}{3pt}
				\centering
				\begin{tabular}{| c | c | c |}
					\hline
					& Real Parameters $\{q, q'\}$ & Complex Parameters $\{z, \bar{z}\}$\\
					\hline
					Evolution Generators & \begin{tabular}{l}
						$0 = \partial_t K_q - \partial_q H + \i \left[H, K_q\right]$\\
						$0 = \partial_t K_{q'} - \partial_{q'} H + \i \left[H, K_{q'}\right]$\\
						$0 = \partial_q K_{q'} - \partial_{q'} K_q + \i \left[K_q, K_{q'}\right]$
					\end{tabular} & \begin{tabular}{l}
						$0 = \partial_t K_z - \partial_z H + \i \left[H, K_z\right]$\\
						$0 = \partial_t K_{\bar{z}} - \partial_{\bar{z}} H + \i \left[H, K_{\bar{z}}\right]$\\
						$0 = \partial_z K_{\bar{z}} - \partial_{\bar{z}} K_z + \i \left[K_z, K_{\bar{z}}\right]$
					\end{tabular}\\
					\hline
					State Evolution & \begin{tabular}{l}
						$0 = \nabla_t \ket{\Psi} = \left(\partial_t + \i H\right) \ket{\Psi}$\\
						$0 = \nabla_q \ket{\Psi} = \left(\partial_q + \i K_q\right) \ket{\Psi}$\\
						$0 = \nabla_{q'} \ket{\Psi} = \left(\partial_{q'} + \i K_{q'}\right) \ket{\Psi}$
					\end{tabular} & \begin{tabular}{l}
						$0 = \nabla_t \ket{\Psi} = \left(\partial_t + \i H\right) \ket{\Psi}$\\
						$0 = \nabla_z \ket{\Psi} = \left(\partial_z + \i K_z\right) \ket{\Psi}$\\
						$0 = \nabla_{\bar{z}} \ket{\Psi} = \left(\partial_{\bar{z}} + \i K_{\bar{z}}\right) \ket{\Psi}$
					\end{tabular}\\
					\hline
					Metric Evolution & \begin{tabular}{l}
						$0 = \nabla_t G = \partial_t G - \i G H + \i H^\dagger G$\\
						$0 = \nabla_q G = \partial_q G - \i G K_q + \i K_q^\dagger G$\\
						$0 = \nabla_{q'} G = \partial_{q'} G - \i G K_{q'} + \i K_{q'}^\dagger G$
					\end{tabular} & \begin{tabular}{l}
						$0 = \nabla_t G = \partial_t G - \i G H + \i H^\dagger G$\\
						$0 = \nabla_z G = \partial_z G - \i G K_z + \i K_{\bar{z}}^\dagger G$\\
						$0 = \nabla_{\bar{z}} G = \partial_{\bar{z}} G - \i G K_{\bar{z}} + \i K_z^\dagger G$
					\end{tabular}\\
					\hline
				\end{tabular}
				\caption{Comparison of evolution generators, state evolutions, and metric evolutions between general real parameters $\{q, q'\}$ and a complex conjugate pair $\{z, \bar{z}\}$. Note that unlike the real-parameter case, metric evolution under complex parameters requires mixed contributions from both generators ($K_z$ and $K_{\bar{z}}$).} \label{Table:Comparison}
			\end{minipage}
		\end{table}

		In summary, the state evolution equations and generators for complex parameters share a structural similarity with their real counterparts. However, unlike the state evolution, the evolution of the metric is not governed by a single generator alone, but is determined by a combination of both the generator and its conjugate counterpart.

	\section{Examples}

		To illustrate the theoretical framework developed in the previous sections, three explicit examples are worked out below.

		As noted in Sec.~\ref{Sec:Review}, parameter generators are not uniquely determined; hence, a gauge choice is required. We therefore adopt the gauge conditions
		\begin{align}
			\left[\partial_t K_z, H\right] = 0 \quad \text{and} \quad \left[\partial_t K_{\bar{z}}, H\right] = 0,
		\end{align}
		which renders both $K_z$ and $K_{\bar{z}}$ at most linear in $t$, i.e.,
		\begin{align}
			K_z = t \p{K}{1}_z + \p{K}{0}_z \quad \text{and} \quad K_{\bar{z}} = t \p{K}{1}_{\bar{z}} + \p{K}{0}_{\bar{z}}.
		\end{align}
		In other words, the gauge conditions reduce to
		\begin{align}
			& \left[\p{K}{1}_z, H\right] = 0,\\
			& \left[\p{K}{1}_{\bar{z}}, H\right] = 0,
		\end{align}
		while Eqs.~\eqref{KZEq} and \eqref{KZBarEq} simplify to the algebraic equations
		\begin{align}
			& \p{K}{1}_z - \partial_z H + \i \left[H, \p{K}{0}_z\right] = 0,\\
			& \p{K}{1}_{\bar{z}} - \partial_{\bar{z}} H + \i \left[H, \p{K}{0}_{\bar{z}}\right] = 0,
		\end{align}
		and Eq.~\eqref{KZandKZBarEq} becomes
		\begin{align}
			& \partial_z \p{K}{1}_{\bar{z}} - \partial_{\bar{z}} \p{K}{1}_z + \i \left[\p{K}{1}_z, \p{K}{0}_{\bar{z}}\right] + \i \left[\p{K}{0}_z, \p{K}{10}_{\bar{z}}\right] = 0,\\
			& \partial_z \p{K}{0}_{\bar{z}} - \partial_{\bar{z}} \p{K}{0}_z + \i \left[\p{K}{0}_z, \p{K}{0}_{\bar{z}}\right] = 0.
		\end{align}

		Besides computational convenience, the physical motivations underlying this gauge choice are detailed in \cite{Ju2024, Ju2024a, Ju2025a}.

		For simplicity and demonstration purposes, the examples in this section will be evaluated along a path $\mathcal{C}$ parameterized by $\theta$, defined by $t = 0$ and $z = e^{\i \theta}$ (and consequently, $\bar{z} = e^{- \i \theta}$). Moreover, since all the Hamiltonian in the following examples are Hermitian at $\theta = 0$, it naturally follows that $G(0) = \mathds{1}$.

		\subsection{Hermitian Case \label{Sec:HermitianExample}}

			This subsection considers the Hamiltonian~\cite{Berry1984}
			\begin{align}
				H = \begin{pmatrix}
					0 & \bar{z}\\
					z & 0
				\end{pmatrix}.
			\end{align}

			For this system, the parameter evolution generators are given by
			\begin{align}
				K_z & = \frac{1}{4 z} \begin{pmatrix}
					- \i & 2 t \bar{z}\\
					2 t z & \i
				\end{pmatrix},\\
				K_{\bar{z}} & = \frac{1}{4 \i \bar{z}} \begin{pmatrix}
					\i & 2 t \bar{z}\\
					2 t z & - \i
				\end{pmatrix}.
			\end{align}

			To track how a general state evolves along the path $C$, the governing evolution equation is
			\begin{align}
				\dv{}{\theta} \ket{\Psi_{\mathcal{C}}} & = \left[\left(\dv{t}{\theta} \partial_t + \dv{z}{\theta} \partial_z + \dv{\bar{z}}{\theta} \partial_{\bar{z}}\right) \ket{\Psi}\right]_{\mathcal{C}}\\
				& = \left[\left(\i z \partial_z - \i \bar{z} \partial_{\bar{z}}\right) \ket{\Psi}\right]_{\mathcal{C}}\\
				& = \left(z K_z - \bar{z} K_{\bar{z}}\right)_{\mathcal{C}} \ket{\Psi_{\mathcal{C}}},
			\end{align}
			where the subscript $\mathcal{C}$ denotes evaluation along the path $\mathcal{C}$, and
			\begin{align}
				\left(z K_z - \bar{z} K_{\bar{z}}\right)_{\mathcal{C}} = \frac{1}{2} \begin{pmatrix}
					1 & 0\\
					0 & -1
				\end{pmatrix}.
			\end{align}

			A straightforward integration yields the state vector
			\begin{align}
				\ket{\Psi_{\mathcal{C}}} = \begin{pmatrix}
					\alpha e^{- \i \theta / 2}\\
					\beta e^{\i \theta / 2}
				\end{pmatrix},
			\end{align}
			where $\alpha$ and $\beta$ are arbitrary constants.

			On the other hand, the metric tensor $G$, governed by
			\begin{align}
				\dv{}{\theta} G_{\mathcal{C}} & = \left[\left(\dv{t}{\theta} \partial_t + \dv{z}{\theta} \partial_z + \dv{\bar{z}}{\theta} \partial_{\bar{z}}\right) G \right]_{\mathcal{C}}\\
				& = \left[\left(\i z \partial_z - \i \bar{z} \partial_{\bar{z}}\right) G\right]_{\mathcal{C}}\\
				& = \left[z \left(- G_{\mathcal{C}} K_z + K_{\bar{z}}^\dagger G_{\mathcal{C}}\right) - \bar{z} \left(- G_{\mathcal{C}} K_{\bar{z}} + K_z^\dagger G_{\mathcal{C}}]\right)\right]_\mathcal{C}\\
				& = \left(z K_{\bar{z}}^\dagger - \bar{z} K_z^\dagger\right)_{\mathcal{C}} G_{\mathcal{C}} + G_{\mathcal{C}} \left(- z K_z + \bar{z} K_{\bar{z}}\right)_{\mathcal{C}},
			\end{align}
			where
			\begin{align*}
				\left(z K_{\bar{z}}^\dagger - \bar{z} K_z^\dagger\right)_{\mathcal{C}} = \frac{1}{2} \begin{pmatrix}
					1 & 0\\
					0 & -1
				\end{pmatrix},
			\end{align*}
			leads to the solution
			\begin{align}
				G_{\mathcal{C}} = \begin{pmatrix}
					1 & 0\\
					0 & 1
				\end{pmatrix},
			\end{align}
			for any $\theta$, given $G(0) = \mathds{1}$ as introduced at the beginning of this section. This result is expected since the Hamiltonian is Hermitian for any given $\theta$ along the path.

			Finally, the state normalization is examined via
			\begin{align}
				\Braket{\Psi_{\mathcal{C}}}{\Psi_{\mathcal{C}}} & = \bra{\Psi_{\mathcal{C}}} G_{\mathcal{C}} \ket{\Psi_{\mathcal{C}}}\\
				& = \begin{pmatrix}
					\bar{\alpha} e^{\i \theta / 2} & \bar{\beta} e^{- \i \theta / 2}
				\end{pmatrix} \begin{pmatrix}
					1 & 0\\
					0 & 1
				\end{pmatrix} \begin{pmatrix}
					\alpha e^{- \i \theta / 2}\\
					\beta e^{\i \theta / 2}
				\end{pmatrix}\\
				& = \alpha \bar{\alpha} + \beta \bar{\beta}.
			\end{align}
			As shown, the $\theta$-dependence completely drops out of the squared norm. Consequently, if the state is normalized such that $\alpha \bar{\alpha} + \beta \bar{\beta} = 1$ at $\theta = 0$, it remains properly normalized along the entire path $\mathcal{C}$.

		\subsection{$\cal{PT}$-Symmetric Case}

			Next, we turn to a non-Hermitian $\mathcal{PT}$-symmetric (parity-time symmetric) Hamiltonian~\cite{Bender2004},
			\begin{align}
				H = \begin{pmatrix}
					z & 2\\
					2 & \bar{z}
				\end{pmatrix}.
			\end{align}
			To keep the discussion simple and avoid hitting an exceptional point, the off-diagonal elements are set to $2$ rather than $1$ (which might otherwise seem more natural). Detailed discussions regarding the evolution generators at exceptional points can be found in \cite{Ju2024a, Ju2025, Ju2025a}.

			The corresponding $z$- and $\bar{z}$-evolution generators are given by
			\begin{align}
				K_z & = \frac{1}{(z - \bar{z} + 4 \i) (z - \bar{z} - 4 \i)} \begin{pmatrix}
					\left[\left(z - \bar{z}\right)^2 + 8\right] t & 2 \left(z - \bar{z}\right) t + 2 \i\\
					2 \left(z - \bar{z}\right) t - 2 \i & 8 t
				\end{pmatrix},\\
				K_{\bar{z}} & = \frac{1}{(z - \bar{z} + 4 \i) (z - \bar{z} - 4 \i)} \begin{pmatrix}
					8 t & - 2 \left(z - \bar{z}\right) t - 2 \i \bar{z}\\
					- 2 \left(z - \bar{z}\right) t + 2 \i \bar{z} & \left[\left(z - \bar{z}\right)^2 + 8\right] t
				\end{pmatrix}.
			\end{align}

			From the governing evolution equation, namely,
			\begin{align}
				\dv{}{\theta} \ket{\Psi_{\mathcal{C}}} = \left(z K_z - \bar{z} K_{\bar{z}}\right)_{\mathcal{C}} \ket{\Psi_{\mathcal{C}}},
			\end{align}
			where
			\begin{align}
				\left(z K_z - \bar{z} K_{\bar{z}}\right)_{\mathcal{C}} = \frac{2 \cos \theta}{7 + \cos \theta} \begin{pmatrix}
					0 & - 1\\
					1 & 0
				\end{pmatrix},
			\end{align}
			together with a straightforward diagonalization, the general expression for the state vector is found to be
			\begin{align}
				\ket{\Psi_{\mathcal{C}}} = \frac{1}{\sqrt{2}}\begin{pmatrix}
					\alpha \dfrac{\left(2 + \sin \theta\right)^{1/4}}{\left(2 - \sin \theta\right)^{1/4}} + \beta \dfrac{- \i \left(2 - \sin \theta\right)^{1/4}}{\left(2 + \sin \theta\right)^{1/4}}\\
					\beta \dfrac{\left(2 + \sin \theta\right)^{1/4}}{\left(2 - \sin \theta\right)^{1/4}} - \alpha \dfrac{- \i \left(2 - \sin \theta\right)^{1/4}}{\left(2 + \sin \theta\right)^{1/4}}
				\end{pmatrix},
			\end{align}
			where $\alpha$ and $\beta$ are arbitrary constants.

			As for the metric tensor $G$, the governing equation is
			\begin{align}
				\dv{}{\theta} G_{\mathcal{C}} = \left(z K_{\bar{z}}^\dagger - \bar{z} K_z^\dagger\right)_{\mathcal{C}} G_{\mathcal{C}} + G_{\mathcal{C}} \left(- z K_z + \bar{z} K_{\bar{z}}\right)_{\mathcal{C}},
			\end{align}
			where
			\begin{align*}
				\left(z K_{\bar{z}}^\dagger - \bar{z} K_z^\dagger\right)_{\mathcal{C}} = \frac{2 \cos \theta}{7 + \cos \theta} \begin{pmatrix}
					0 & 1\\
					- 1 & 0
				\end{pmatrix} = - \left(z K_z - \bar{z} K_{\bar{z}}\right)_{\mathcal{C}}.
			\end{align*}
			Therefore, with the initial condition $G(0) = \mathds{1}$, the corresponding metric $G_{\mathcal{C}}$ reads
			\begin{align}
				G_{\mathcal{C}} = \frac{1}{\left(4 - \sin^2 \theta\right)^{1/2}} \begin{pmatrix}
					2 & - \i \sin \theta\\
					\i \sin \theta & 2
				\end{pmatrix}.
			\end{align}

			Finally, a nontrivial check of the state normalization yields
			\begin{align}
				\Braket{\Psi_{\mathcal{C}}}{\Psi_{\mathcal{C}}} & = \bra{\Psi_{\mathcal{C}}} G_{\mathcal{C}} \ket{\Psi_{\mathcal{C}}} = \alpha \bar{\alpha} + \beta \bar{\beta}.
			\end{align}
			As in the Hermitian example shown in Sec.~\ref{Sec:HermitianExample}, the $\theta$-dependence completely drops out of the normalization expression. Consequently, the inner product and physical norm are perfectly preserved along the entire path $\mathcal{C}$.

		\subsection{$\bar{z}$-Independent Case}

			Finally, we demonstrate the case where the Hamiltonian depends solely on $z$ rather than $\bar{z}$~\cite{MehriDehnavi2008}, namely,
			\begin{align}
				H = \begin{pmatrix}
					0 & 1\\
					z & 0
				\end{pmatrix}.
			\end{align}

			The corresponding $z$- and $\bar{z}$-evolution generators obtained from the Hamiltonian are given by
			\begin{align}
				K_z & = \frac{1}{4 z} \begin{pmatrix}
					- \i & 2 t\\
					2 z t & \i
				\end{pmatrix},\\
				K_{\bar{z}} & = 0,
			\end{align}
			where the vanishing $\bar{z}$-evolution generator is a direct consequence of the $\bar{z}$-independence of $H$. 

			The state evolution is then governed by
			\begin{align}
				\dv{}{\theta} \ket{\Psi_{\mathcal{C}}} = \left(z K_z\right)_{\mathcal{C}} \ket{\Psi_{\mathcal{C}}},
			\end{align}
			which leads to the general solution
			\begin{align}
				\ket{\Psi_{\mathcal{C}}} = \begin{pmatrix}
					\alpha e^{- \i \theta / 4}\\
					\beta e^{\i \theta / 4}
				\end{pmatrix},
			\end{align}
			where $\alpha$ and $\beta$ are arbitrary constants.

			Furthermore, the metric tensor $G_{\mathcal{C}}$ is determined by the governing equation
			\begin{align}
				\dv{}{\theta} G_{\mathcal{C}} = - \left(\bar{z} K_z^\dagger\right)_{\mathcal{C}} G_{\mathcal{C}} - G_{\mathcal{C}} \left(z K_z\right)_{\mathcal{C}},
			\end{align}
			which renders the solution
			\begin{align}
				G_{\mathcal{C}} = \begin{pmatrix}
					1 & 0\\
					0 & 1
				\end{pmatrix},
			\end{align}
			for any $\theta$.

			Finally, the state norm is evaluated as
			\begin{align}
				\Braket{\Psi_{\mathcal{C}}}{\Psi_{\mathcal{C}}} & = \bra{\Psi_{\mathcal{C}}} G_{\mathcal{C}} \ket{\Psi_{\mathcal{C}}} = \alpha \bar{\alpha} + \beta \bar{\beta}.
			\end{align}
			As always, once the state is normalized at the initial boundary, it remains properly normalized for all $\theta$ along the path $\mathcal{C}$.

	\section{Conclusions}

		Variations in a quantum system can frequently be described as changes in the parameters of its corresponding Hamiltonian. Consequently, analyzing how a system responds to external factors requires understanding its dependence on Hamiltonian parameters, a task traditionally approached using Berry connections or Berry phases.

		Unlike a Berry connection, which tracks the evolution of a specific eigenstate or a subset of eigenstates rather than the full Hilbert space, recent developments establish a systematic framework for finding connections over the entire Hilbert space bundle by simply enforcing normalization preservation. However, this framework has historically been restricted to real parameters, often requiring complex quantities to be split into separate real components.

		Because physical parameters such as coupling constants are generally complex even within Hermitian systems, limiting the framework to real parameters imposes an unnecessary constraint.

		To overcome this limitation, complex parameters $z$ are decomposed into real variables to construct a general formalism via Wirtinger derivatives. Specifically, evolution generators in the $z$-direction are obtained by treating a complex parameter and its conjugate as independent variables while adapting the governing equations established in real parameter spaces.

		While the $z$-evolution of quantum states is naturally governed by the corresponding $z$-evolution generator, the evolution of the metric requires a coupled contribution from both the $z$- and $\bar{z}$-generators. Alongside this formal derivation, explicit examples illustrate how the formalism is implemented in practice.

		Ultimately, by properly incorporating Wirtinger derivatives into the parallel transport bundle structure, this extension provides a rigorous and versatile machinery for investigating the geometric and dynamical responses of quantum systems under complex parameter variations.

	\section*{Acknowledgments}
		J.Y.Y and C.Y.J. are partially supported by the National Science and Technology Council under Grant No. NSTC 115-2112-M-110-011.

	\section*{Appendix}
	\begin{appendix}
		\setcounter{equation}{0}
		\renewcommand{\theequation}{\thesection.\arabic{equation}}

		\section{Derivation of the Evolution and Compatibility Conditions for Complex Parameters \label{Appendix:DetailedDerivations}}

			To derive the $z$-evolution for the metric $G$, applying the Wirtinger derivatives in Eq.~\eqref{WirtingerDerivative} renders
			\begin{align}
				\partial_z G & = \frac{1}{2} \left(\partial_x - \i \partial_y\right) G\\
				& = \frac{1}{2}\left(\i G K_x - \i K_x^\dagger G\right) - \frac{\i}{2}\left(\i G K_y - \i K_y^\dagger G\right)\\
				& = \i G \left(\frac{K_x - \i K_y}{2}\right) - \i \left(\frac{K_x^\dagger - \i K_y^\dagger}{2}\right) G\\
				& = \i G \left(\frac{K_x - \i K_y}{2}\right) - \i \left(\frac{K_x + \i K_y}{2}\right)^\dagger G\\
				& = \i G K_z - \i K_{\bar{z}}^\dagger G,
			\end{align}
			which is exactly the evolution equation appearing in Eq.~\eqref{GZEq}. The $\bar{z}$ version for the metric $G$ follows almost identically from the derivation above.

			To examine the compatibility conditions between the Hamiltonian $H$, $K_z$, and $K_{\bar{z}}$, the Wirtinger derivatives in Eq.~\eqref{WirtingerDerivative} and the relations in Eqs.~\eqref{KXYEq1}, \eqref{KXYEq2}, and \eqref{KXYEq3} are utilized. The governing equations for $K_z$ and $K_{\bar{z}}$ are obtained via
			\begin{align}
				& \begin{cases}
					\partial_t K_x = \i \left[K_x, H\right] + \partial_x H,\\
					\partial_t K_y = \i \left[K_y, H\right] + \partial_y H,
				\end{cases}\\
				\Rightarrow & \begin{cases}
					\partial_t K_x = \i \left[K_x, H\right] + \partial_z H + \partial_{\bar{z}} H,\\
					\partial_t K_y = \i \left[K_y, H\right] + \i \partial_{\bar{z}} H - \i \partial_z H,
				\end{cases}\\
				\Rightarrow & \begin{cases}
					\partial_t \left(K_x - \i K_y\right) = \i \left[K_x - \i K_y, H\right] + 2 \partial_z H,\\
					\partial_t \left(K_x + \i K_y\right) = \i \left[K_x + \i K_y, H\right] + 2 \partial_{\bar{z}} H,
				\end{cases}\\
				\Rightarrow & \begin{cases}
					\partial_t K_z = \i \left[K_z, H\right] + \partial_z H,\\
					\partial_t K_{\bar{z}} = \i \left[K_{\bar{z}}, H\right] + \partial_{\bar{z}} H.
				\end{cases}
			\end{align}

			Furthermore, the compatibility condition between $K_z$ and $K_{\bar{z}}$ can be expressed as
			\begin{align}
				& \partial_z K_{\bar{z}} - \partial_{\bar{z}} K_z + \i \left[K_z, K_{\bar{z}}\right]\\
				& = \frac{1}{4} \left(\partial_x - \i \partial_y\right) \left(K_x - \i K_y\right) - \frac{1}{4} \left(\partial_x + \i \partial_y\right) \left(K_x + \i K_y\right) + \i \frac{1}{4} \left[K_x - \i K_y, K_x + \i K_y\right]\\
				& = \frac{- \i}{2} \left(\partial_x K_y + \partial_y K_x + \i \left[K_x, K_y\right]\right) = 0.
			\end{align}

			Therefore, although $z$ and $\bar{z}$ are mutually conjugate, from the perspective of the evolution generators, they can be treated as independent parameters.
	\end{appendix}
	\bibliography{References}
\end{document}